\documentclass[conference]{IEEEtran}
\usepackage{cite}
\usepackage{amsmath,amssymb,amsfonts}
\usepackage{algorithmic}
\usepackage{textcomp}
\usepackage{xcolor}
\usepackage{color}
\usepackage{graphicx} 
\usepackage{subcaption} 
\usepackage{verbatim} 
\def\BibTeX{{\rm B\kern-.05em{\sc i\kern-.025em b}\kern-.08em
    T\kern-.1667em\lower.7ex\hbox{E}\kern-.125emX}}
\begin{document}

\title{An Open-Source Simulation and Data Management Tool for EnergyPlus Building Models
{\footnotesize \textsuperscript{}}
\thanks{}
}

\author{\IEEEauthorblockN{Ninad Gaikwad, Kasey Dettlaff, Athul Jose P and Anamika Dubey} 
\IEEEauthorblockA{\textit{School of Electrical Engineering \& Computer Science} \\
\textit{Washington State University}\\
Pullman, USA \\
ninad.gaikwad@wsu.edu, kasey.dettlaff@wsu.edu, athul.p@wsu.edu and anamika.dubey@wsu.edu}
}

\maketitle

\begin{abstract}
We present a new open-source, GUI-based application created using Plotly-Dash, along with an integrated PostgreSQL-based relational database, developed to streamline EnergyPlus building model simulation workflows. The application facilitates data generation, aggregation (across thermal zones), and visualization based on customizable user preferences, while the database efficiently stores and retrieves complex simulation data generated by EnergyPlus. We demonstrate the need for this application and database, emphasizing how existing approaches for generating, managing, and analyzing EnergyPlus simulation data can be cumbersome, particularly when handling a large number of building models with varying simulation setups. This integrated framework enables building energy engineers and researchers to simplify their EnergyPlus simulations, manage generated simulation data, perform data analyses, and support data-driven modeling tasks.
\end{abstract}


\section{Introduction}\label{sec:Introduction}

Buildings play a pivotal role in energy consumption, accounting for 27.6\% of the total U.S. end-use energy in 2023 across residential and commercial sectors~\cite{EIAEnergyConsumptionWebpage}. Modern buildings, equipped with HVAC systems, battery storage, electric vehicles, and solar photovoltaic (PV) systems, are increasingly seen as critical grid-edge resources, with potential to contribute to energy efficiency and decarbonization.~\cite{xiang2022historical}. Passive strategies, such as retrofitting and energy-efficient planning, reduce carbon emissions while supporting the grid~\cite{zhou2016achieving}. Active measures such as demand response and advanced control strategies enable buildings to help integrate intermittent renewable energy sources while improving grid stability~\cite{chen2018measures}. These capabilities illuminate the essential role of buildings in shaping a sustainable energy future. 

To enable buildings to act as effective grid-edge resources, robust data collection, analysis, and modeling are essential to inform decarbonization strategies. The implementation of technologies such as demand response will require dynamic building models that can interface with intelligent controllers for real-time decision-making~\cite{shan2016building}. In these applications, generating, aggregating, and analyzing large amounts of data is critical~\cite{gunay2019data}. However, limited access to real-world building data remains a major challenge to research efforts. To address this challenge, white-box simulation tools such as EnergyPlus~\cite{EnergyPlusToolsDirectory} are widely used to generate high-fidelity data for these purposes. Additionally, approaches based in machine learning may leverage data from these simulations to develop practical models for use in real-world applications~\cite{alanne2022overview}.


The EnergyPlus Ecosystem offers various tools tailored to specific needs, each with unique applications and limitations. OpenStudio~\cite{OpenStudioDocumentation}, a widely used graphical interface for EnergyPlus, allows users to create and modify input files (IDFs) by designing building geometry, occupancy schedules, lighting, ventilation, and HVAC setpoints. Tools like CityBES~\cite{CityBESDocumentation} enable large-scale modeling using EnergyPlus, while software such as EDAPT, EFEN, and ESP-r cater to whole-building simulations. Numerous other tools exist for applications ranging from HVAC selection to building automation and more. However, these tools include features that are extraneous for users who simply want an improved interface to run simulations for varied scenarios, aggregate this data in a meaningful manner, and perform basic statistical analysis with visualization. Our simulation management tool streamlines this workflow by removing unnecessary complexity introduced by existing tools. 

A storage solution for building simulation data generated from EnergyPlus is needed to store custom simulation data in an efficient and flexible manner. Currently,  ResStock~\cite{present2024resstock} and ComStock~\cite{parker2023comstock} offer large databases of pre-simulated building energy data for residential and commercial properties across the United States. Through their web interfaces, users can explore time-series data for variables such as electricity consumption for cooling, heating, and lighting. These tools only provide a curated selection of data, without the option for custom simulations or variable selection. This underscores the need for a data management tool that allows users to create, manage, and query data from their own custom simulations. Our data management tool offers an efficient and flexible framework for storing large volumes of building data, accommodating any variable selection or building configuration.

The remainder of this paper is organized as follows: Section~\ref{sec:EnergyPlusPNNLDatabase} provides an overview of EnergyPlus and the PNNL Prototypical Buildings Database. Section~\ref{sec:SimManagement} details the Simulation Management Tool, highlighting its architecture and graphical interface. Section~\ref{sec:DataManagement} explains the Data Management Tool, focusing on its relational database schema designed to efficiently store, query, and analyze time-series data. Finally, Section~\ref{sec:Conclusion} concludes the paper by summarizing the key findings and discussing future directions for enhancing simulation and data management in building energy research.

\section{EnergyPlus and PNNL Prototypical Buildings Database}\label{sec:EnergyPlusPNNLDatabase}

\subsection{Description and Limitations of EnergyPlus}\label{subsec:DescriptionandLimitationofEnergyPlus}
\begin{figure}
    \centering
    \includegraphics[width=1\linewidth]{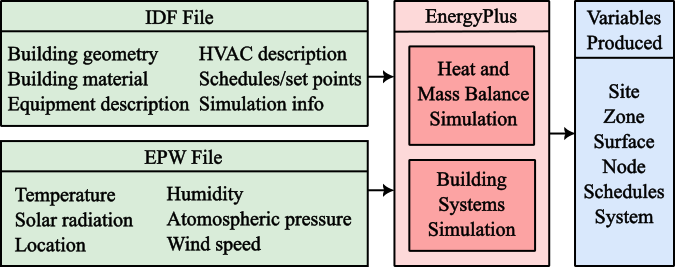}
    \caption{EnergyPlus schematic}
    \label{fig:energyplusschematic}
\end{figure}

EnergyPlus~\cite{EnergyPlus} is an open-source simulation software designed for modeling of building energy consumption across systems such as heating, cooling, ventilation, lighting, and equipment loads.  It is widely used by engineers and architects to optimize building energy performance and efficiency. Using a heat balance-based approach, EnergyPlus simulates a building’s thermal response and energy use with sub-hourly time steps, integrated HVAC modeling, and environmental interaction. Fig.~\ref{fig:energyplusschematic} illustrates the EnergyPlus workflow, showing the inputs, simulation process, and results. The simulation process relies on two key input files: the Input Data File (IDF) and the EnergyPlus Weather File (EPW). The IDF specifies the building’s geometry, construction materials, HVAC systems, usage schedules, and other configurations, while the EPW provides detailed weather data, including temperature, solar radiation, humidity, and wind speed. Energyplus then calculates the building's thermal loads and equipment energy use, and generates time series data for specified variables. 

Output data is provided in tabular format for a variety of variables, categorized by zone, surface, system node, schedule, and site. Facility-level variables, like total HVAC demand, are represented in a single column, while more detailed variables, such as zone temperature, have separate time-series columns for each attribute. Schedules correspond to occupancy, lighting levels, and equipment levels for up to four kinds of equipment: electric, gas, hot water, and steam. 

EnergyPlus alone has limitations that can impact its usability and accessibility. The software relies on text-based input and output files, which requires users to manually create, process, and interpret data. The need for background knowledge of the file structure and syntax poses a challenge for users without technical expertise or experience using the tool. Furthermore, the lack of built-in options for data visualization and analysis complicates the interpretation of simulation outputs, especially for large or complex projects. Overall, the tool's learning curve limits its accessibility to a wider audience of users.

\subsection{Description of PNNL Prototypical Buildings Data}\label{subsec:PNNLPrototypical Data}
EnergyPlus provides pre-configured IDF files for prototypical buildings, including 5,168 commercial, 3,552 residential, and 152 manufactured buildings. Each prototype is tailored with specific attributes, such as heating systems and foundation types for residential buildings. 

\section{Simulation Management Tool}\label{sec:SimManagement}
\begin{figure}[t]
    \centering
    \includegraphics[width=1\linewidth]{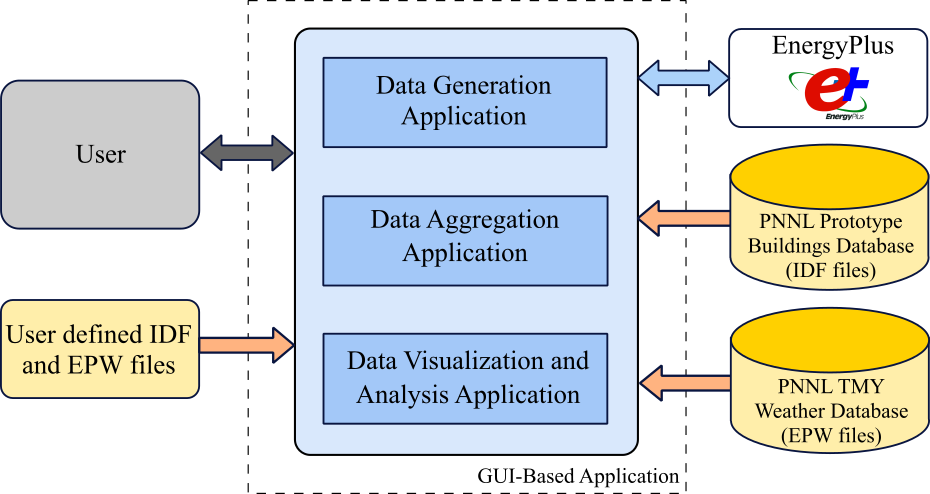}
    \caption{Simulation management tool schematic}
    \label{fig:app_schematic}
\end{figure}

We have developed an open-source, GUI-based simulation management tool to streamline the processes of generating, aggregating, visualizing data from EnergyPlus simulations using Python and the OpyPlus library. Fig.~\ref{fig:app_schematic} shows the high-level design of the tool, which provides a user-friendly graphical interface to interact with EnergyPlus. Users can upload their own IDF files or access the PNNL database to use prototypical models. They can also modify the simulation setup directly within the GUI, avoiding the need to edit IDF files by hand. The tool consists of three modules: the Data Generation Application, Data Aggregation Application, and Data Visualization and Analysis Application, each designed to simplify its respective stage of the simulation workflow.

\begin{figure*}[h!]
    \centering
    \begin{subfigure}{0.32\textwidth}
        \includegraphics[width=\textwidth]{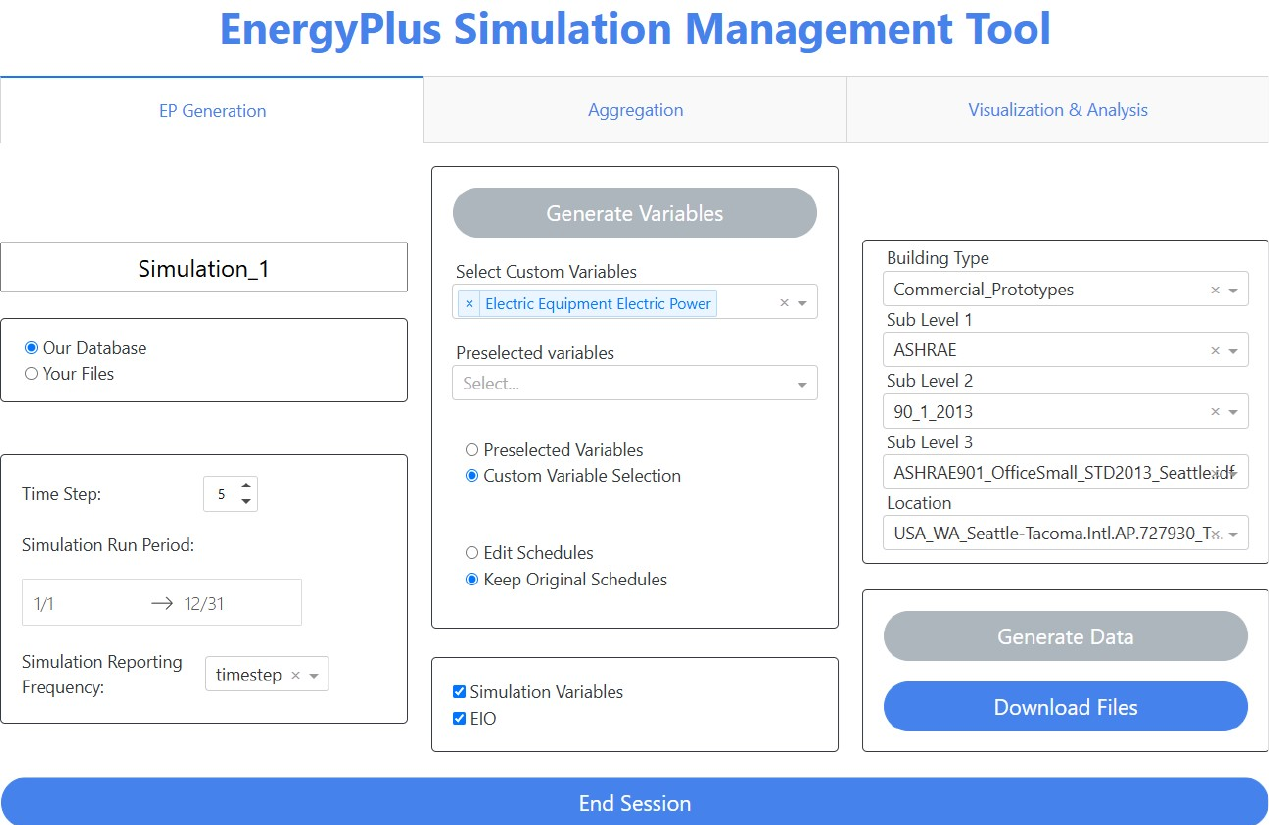}
        \caption{The Data Generation App}
        \label{fig:data_gen_app}
    \end{subfigure}
    \hfill
    \begin{subfigure}{0.32\textwidth}
        \includegraphics[width=\textwidth]{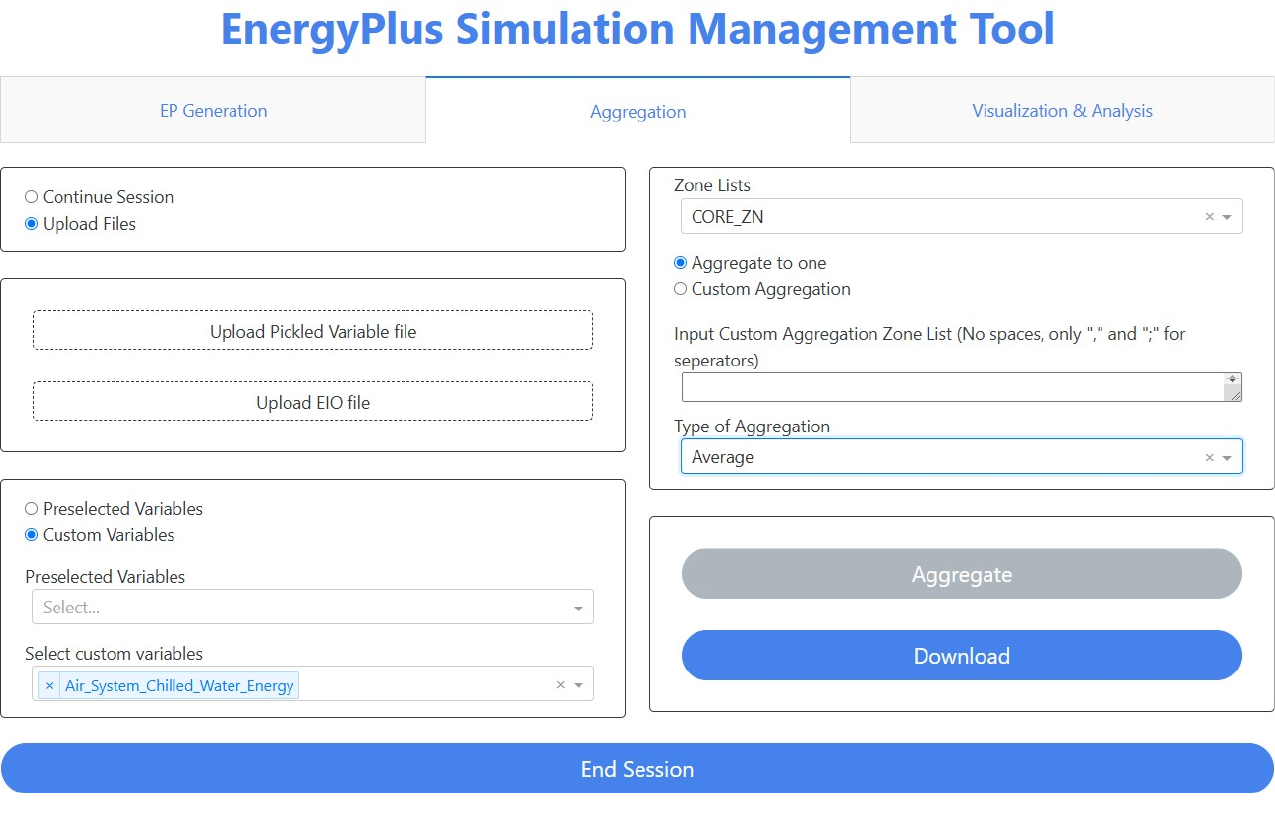}
        \caption{The Data Aggregation App}
        \label{fig:data_agg_app}
    \end{subfigure}
    \hfill
    \begin{subfigure}{0.32\textwidth}
        \includegraphics[width=\textwidth]{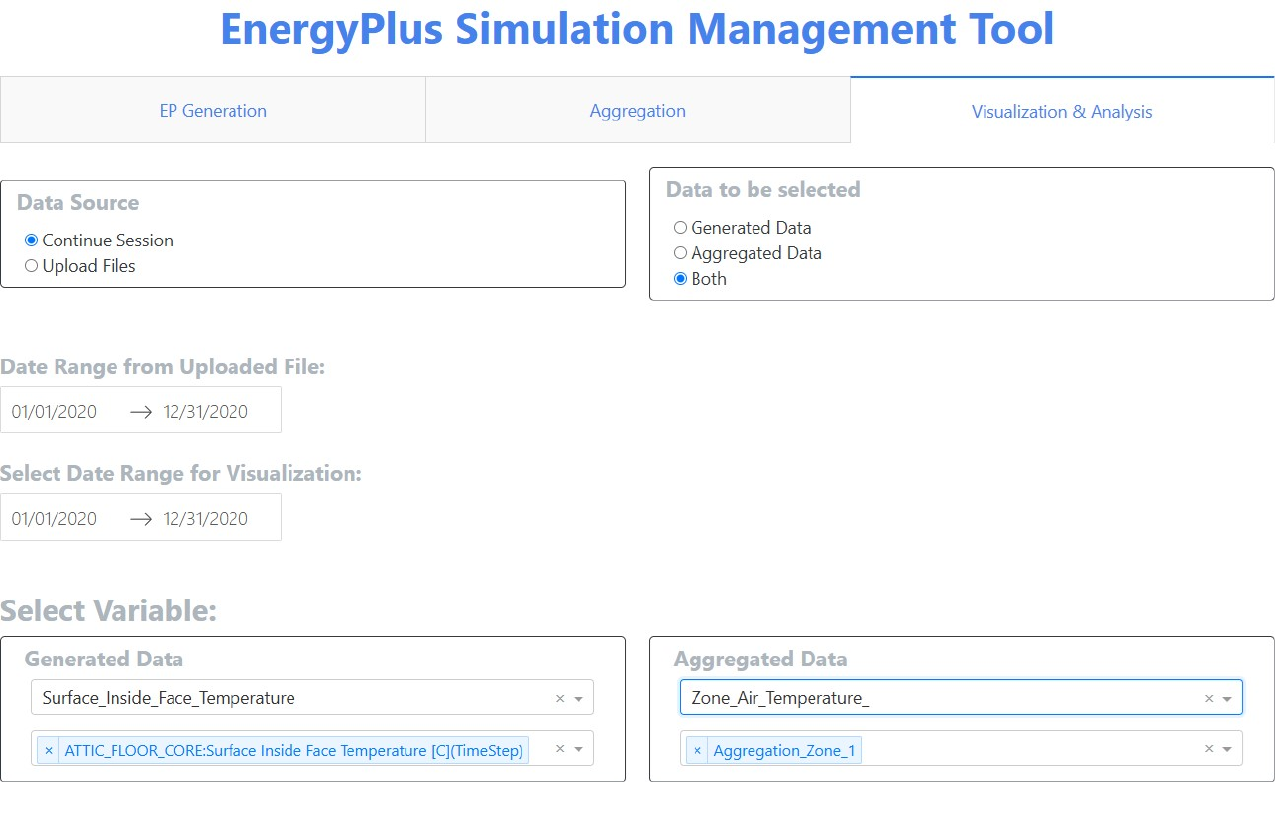}
        \caption{The Data Visualization and Analysis App}
        \label{fig:data_vis_app}
    \end{subfigure}
    \caption{Graphical User Interface of the Apps: Generation, Aggregation, and Visualization.}
    \label{fig:apps_gui}
\end{figure*}

\subsection{Data Generation Application}\label{subsec:DataGenApp}
The Data Generation Application creates data by customizing EnergyPlus simulations according to user specifications. The Data Generation Application, depicted in Fig.~\ref{fig:apps_gui}\subref{fig:data_gen_app}, prompts the user to select between PNNL prototype buildings or upload custom files. For database selection, users navigate through available IDF and weather files. In the simulation configuration module, users adjust settings such as timestep, run period, and reporting frequency. The Data Generation Application performs an initial run of EnergyPlus, revealing all available simulation variables for the given IDF. The user can select from these variables, or choose our pre-selected variable set, consisting of 35 variables key to estimating building power consumption. Schedule customization allows users to input custom schedules for occupants and equipment. After configuring these options, users press the Generate Data button to run the simulation, then press the Download Data button to obtain the generated files. Data is generated in a structured pickle format, \texttt{\{'Var 1': pd.DataFrame, 'Var2': pd.DataFrame, ..., 'VarN': pd.DataFrame \}} with each variable stored as a pandas DataFrame.

\subsection{Data Aggregation Application}\label{subsec:DataAggApp}
The Data Aggregation Application allows users to felxibily combine thermal zones to create datasets optimized for reduced-order thermal modeling. Fig.~\ref{fig:apps_gui}\subref{fig:data_agg_app} shows the GUI of the Data Aggregation Application. Users can upload new files (pickle and EIO), or use previously generated files from the Data Generation Application. After uploading, users select variables for aggregation, either from predefined essential variables or custom-selected ones. Next, users can customize aggregations settings, grouping data into a single zone or multiple custom zones, with the user specifying a list of custom aggregation zones if needed. Users can choose among three aggregation methods—simple average, floor area-weighted average, and volume-weighted average. Once the configuration is complete, selecting the "Aggregate" button initiates the aggregation process, which generates a downloadable output in a structured pickle format. This output organizes the data as \texttt{\{'AggZone1': \{'Var1': pd.DataFrame, 'Var2': pd.DataFrame, ..., 'VarN': pd.DataFrame \}, ..., 'AggZoneM': \{'Var1': pd.DataFrame, 'Var2': pd.DataFrame, ..., 'VarN': pd.DataFrame \}\}}, where each aggregated zone contains pandas DataFrames for the specified variables.

\subsection{Data Visualization and Analysis Application}\label{subsec:DataVisApp}

The Data Visualization and Analysis Application illustrated in Fig.~\ref{fig:apps_gui}\subref{fig:data_vis_app} offers an interface for visualizing and exploring building performance data from the Data Generation and Aggregation Applications or user-uploaded pickle files. Users can generate time-series  plots for any variable over a selected duration, create scatter plots to examine relationships between variables, and display distribution plots with essential statistics, including histograms, means, variances, and ranges. This environment enables clear visualization of time-based trends, variable correlations, and statistics.

\begin{figure}[h!]
    \centering
    \includegraphics[width=\columnwidth]{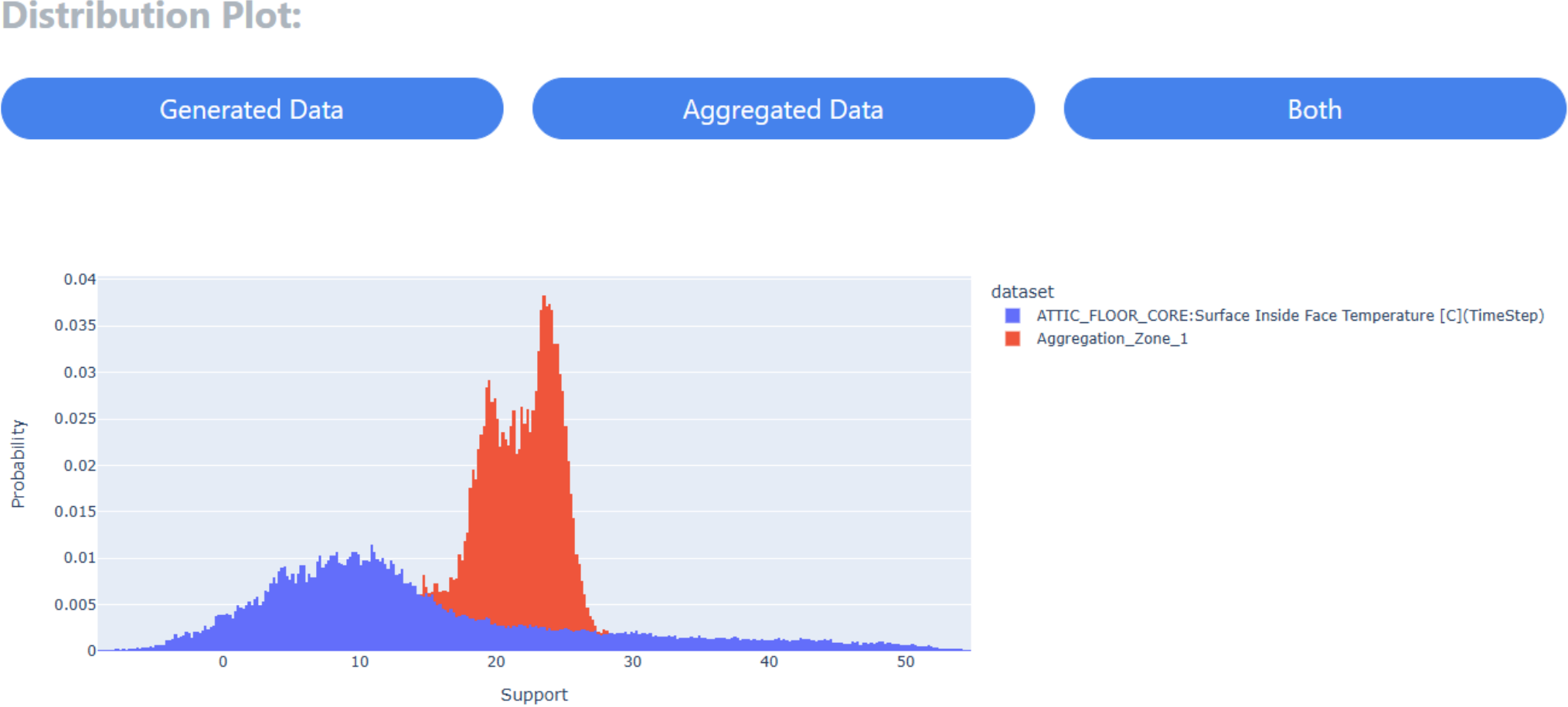}
    \caption{Distribution Plot}
    \label{fig:vis_dist_plot}
\end{figure}

\begin{figure}[h!]
    \centering
    \includegraphics[width=\columnwidth]{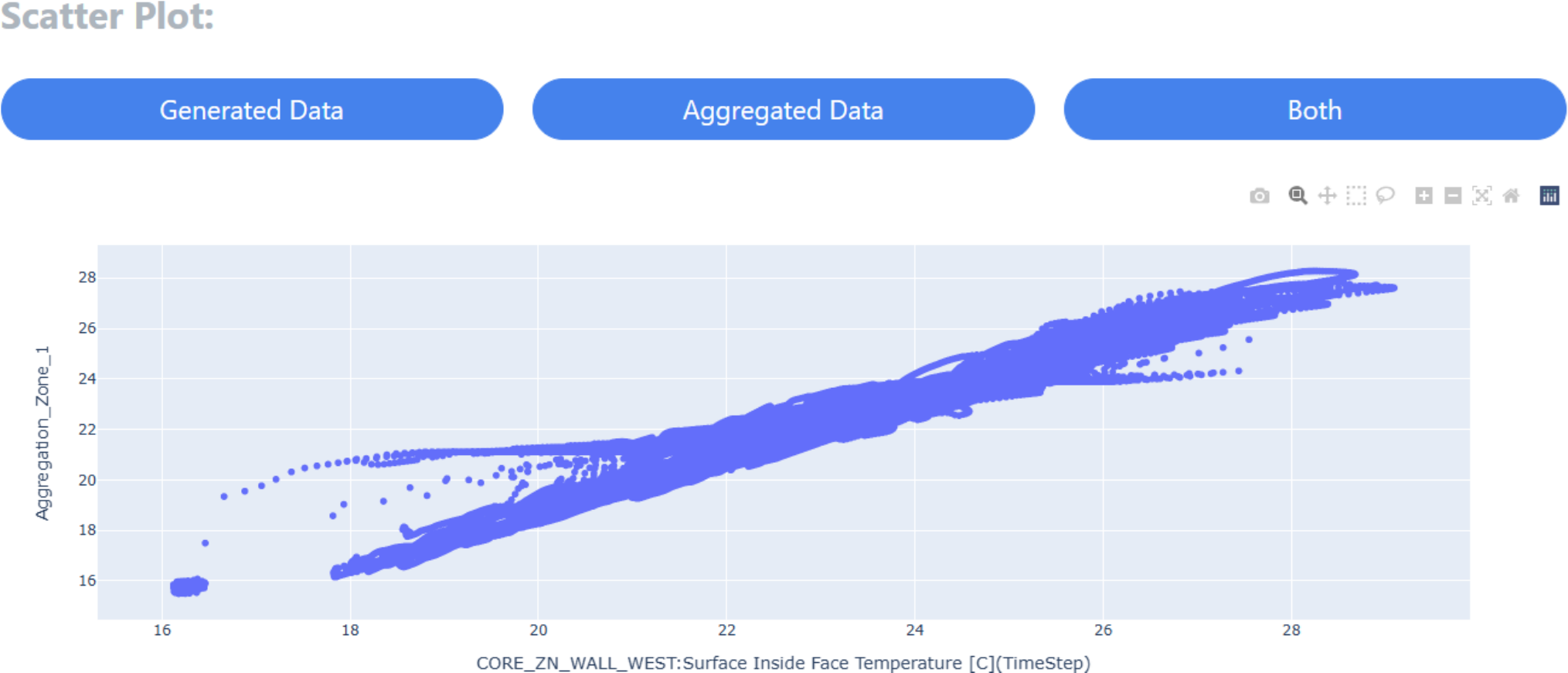}
    \caption{Scatter Plot}
    \label{fig:vis_scat_plot}
\end{figure}

\begin{figure}[h!]
    \centering
    \includegraphics[width=\columnwidth]{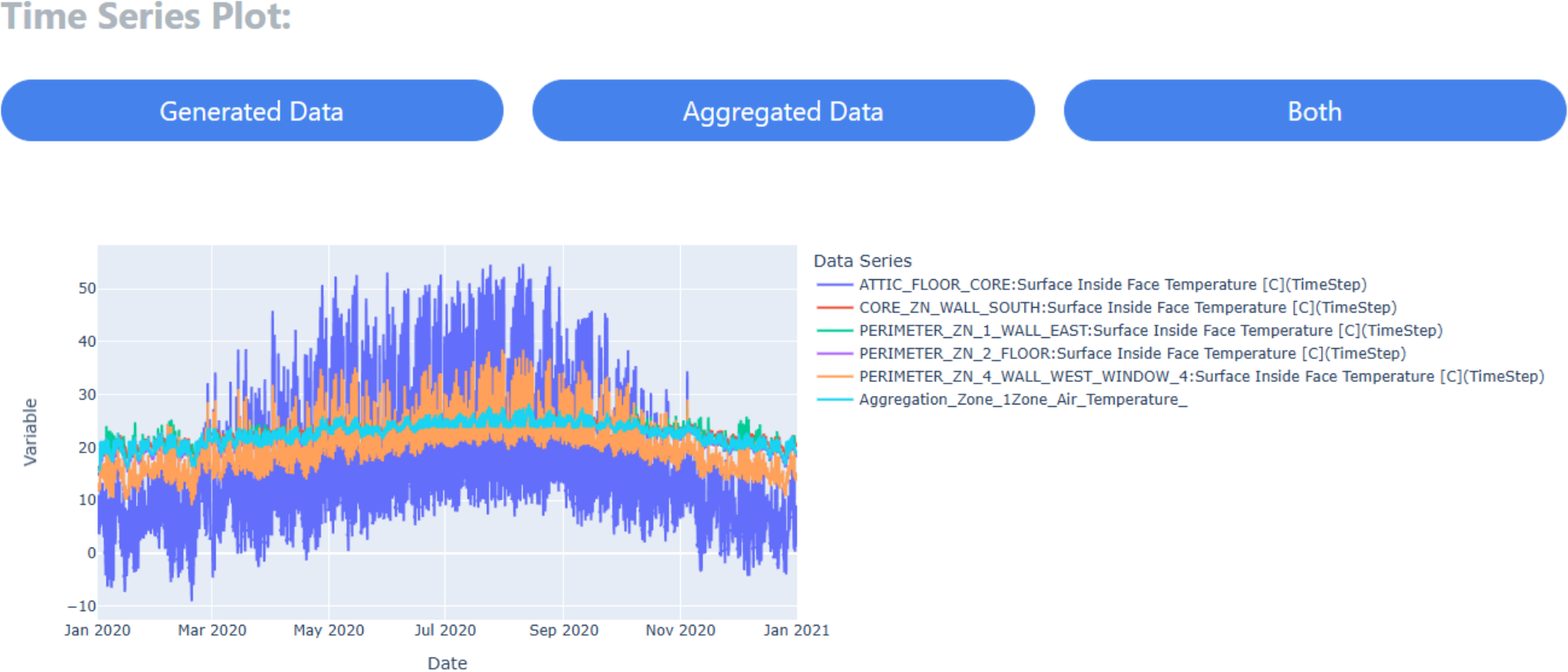}
    \caption{Time Series Plot}
    \label{fig:vis_time_plot}
\end{figure}

Users begin by choosing either to continue from a previous session or to upload the generated and aggregated data files. The Application then prompts them to choose a specific date range for analysis. It offers three primary plot types: a distribution plot (Fig.~\ref{fig:vis_dist_plot}) for single-variable analysis , a scatter plot (Fig.~\ref{fig:vis_scat_plot}) to explore relationships between two selected variables, and a time-series plot (Fig.~\ref{fig:vis_time_plot}) that supports multiple variables to observe trends over time. All plots are interactive, with zoom and pan capabilities enabled through Plotly Dash, and each can be downloaded as a PNG for easy reporting and sharing.

\begin{figure}
    \centering
    \includegraphics[width=\linewidth]{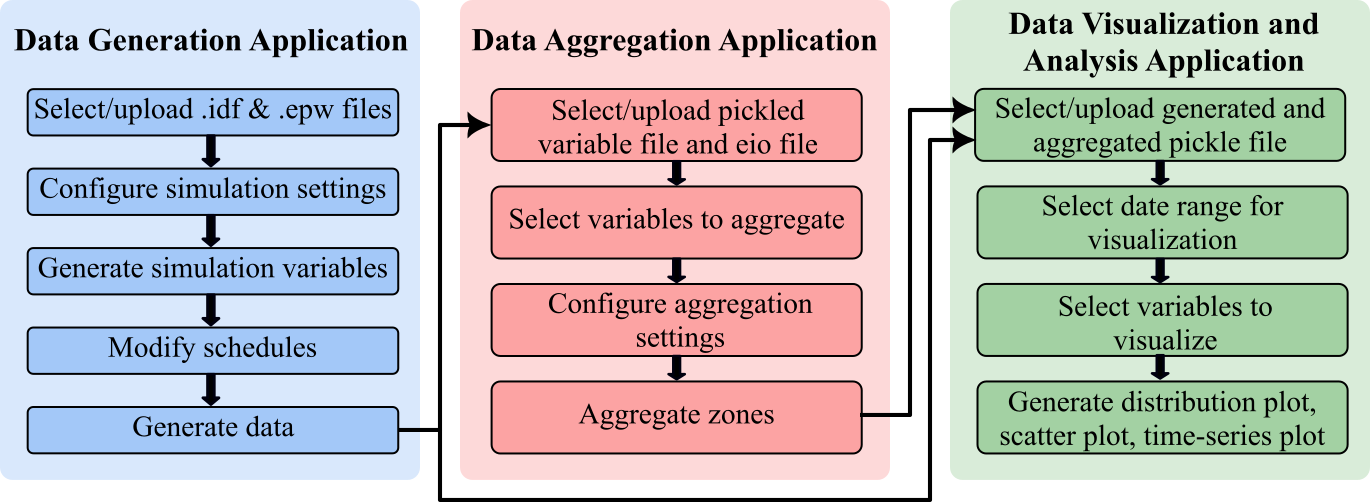}
    \caption{Simulation management tool workflow}
    \label{fig:app_workflow}
\end{figure}

The workflow of the application is illustrated in Fig.~\ref{fig:app_workflow}. Each application is interconnected, allowing users to generate, download, and save data in each session, which can then be uploaded and used in subsequent sessions. Alternatively, users can perform data generation, aggregation, and visualization sequentially without needing to save data between steps.

\section{Data Management Tool}\label{sec:DataManagement}

\begin{figure}
    \centering
    \includegraphics[width=\linewidth]{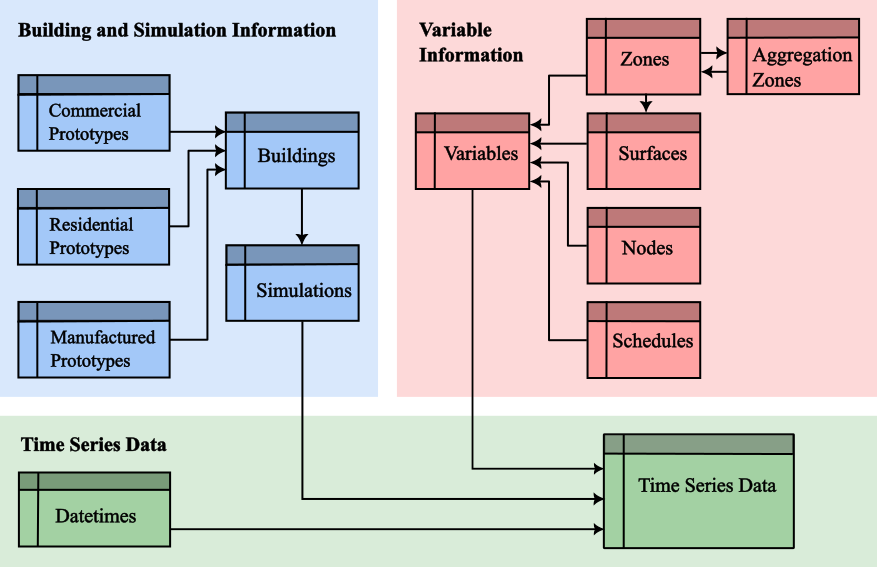}
    \caption{Database Schema}
    \label{fig:database_schema}
\end{figure}


A relational database is a structured data collection organized into linked tables. Each table has a primary key—a unique identifier for each record—ensuring distinct entries. We use integers as primary keys in order to minimize storage and speed up indexing and queries. A foreign key is a field that links one table to another by referencing the primary key, creating relationships between tables. For example, the Time Series Data table references Simulations, Variables, and Datetimes by storing their integer identifiers as foreign keys in respective columns. 

In a database, a linking table is used to manage many-to-many relationships between two tables. This type of relationship occurs when multiple records in one table relate to multiple records in another. For example, consider thermal zones in buildings: each building has multiple thermal zones, and identical zones (with the same name and dimensions) may appear in different buildings across a large dataset. To manage this, we create a linking table between the Zones and Buildings tables, containing only foreign keys, one from each of the related tables. Each record in the linking table represents a connection between a zone and a building.

We've created a database schema aimed at optimizing the storage of large-scale time-series data generated from EnergyPlus simulations. The proposed schema includes separate tables for building information, simulation settings, variable definitions, and datetimes. This design leverages SQL-based compression techniques to address the inefficiencies associated with the pickle file structure, the most important being data redundancy. We achieve this by designing the schema in a way that reduces the storage required for each time series record.

\subsection{Building and Simulation Information}\label{Building and Simulation Information}

The Simulations table, illustrated in blue in Fig.~\ref{fig:database_schema}, stores metadata for each simulation, including simulation id (primary key), building id (foreign key), weather file location, time resolution, and schedule name. The Buildings table, with the primary key building id, contains information such as the climate location and energy standard of the IDF file. Each building prototype type—commercial, residential, and manufactured—is stored in a separate table to account for unique attributes. The Buildings table links to the appropriate prototype table via linking tables. Each record in the buildings table represents a unique IDF file, and each record in the Simulations table represents a unique Energyplus Simulation.

For example, to upload Seattle Small Office, we start by adding the small office prototype to the Commercial Prototypes table. Next, we create a record in the Buildings table with the ASHRAE 2013 energy standard and Seattle's climate zone. Finally, we add a record to the Simulations table with the Seattle weather file location and a 5-minute time resolution.

\subsection{Variable Information}\label{Variable Information}

Time series data from EnergyPlus is stored in tabular format by variable. Some variables generate multiple columns, such as separate time series columns for each zone within a building. Illustrated in red in Fig.~\ref{fig:database_schema}, the Variables table organizes this data with a primary key variable id, and columns for variable name and variable type. Variables can be categorized as schedule-based, node-based, surface-based, or zone-based. To manage zone-based variables, an additional table handles data aggregation. Aggregated zones, which combine multiple composite zones (original zones from the IDF file), are stored as entries in the Zones table. The Aggregation Zones table defines the many-to-many relationships between aggregated and composite zones.

Continuing the earlier example, we select 35 variables spanning all variable types. The Zones table is populated with entries for each of the five zones in the small office building. These zones are then aggregated into a single zone which is added to the Zones table, with composite zone details stored in the Aggregation Zones table. Data from schedules, nodes, and surfaces are recorded in their respective tables, while the name and type of each selected variable is logged in the Variables table.

\subsection{Time Series Data}\label{Time Series Data}
The Time Series Data table, illustrated in green in Fig.~\ref{fig:database_schema}, is optimized to minimize storage required for each record. It has four columns, three of which are primary keys, referencing the Simulation, Variables, and Datetimes table. The final column stores the specific value of a variable for a given simulation at a particular time. To complete the example, the Time Series Data table is populated with data linked to the corresponding simulation, variable, and datetime.

\subsection{Results for Data Management Tool}\label{subsec:ResultsDataManagement}

Our goal is to generate data for all prototypical buildings, for 35 selected variables, at a 5-minute time resolution. Given the schema and known SQL storage requirements, we calculate the estimated storage requirement for an average building to be 1.7 GB, totaling approximately 15 TB for all prototypical buildings. We have simulated a number of commercial buildings and stored the data in pickle file format. Analysis shows that the average commercial building requires approximately 38 GB of storage, meaning the database reduces storage needs by over 20 times. Using a database structure also allows us to manage the complex relationships between buildings, variables, and zones. Overall, we conclude that the proposed SQL database design provides an optimal solution for managing large-scale EnergyPlus building simulation data. 

\section{Conclusion}\label{sec:Conclusion}

This paper introduces an open-source simulation management and data management tool designed to enhance EnergyPlus workflows. The simulation management tool simplifies data generation, aggregation, and visualization, while the data management tool optimizes storage efficiency and usability, enabling more effective data analysis. Together, these tools empower researchers to streamline EnergyPlus simulations, manage simulation data effectively, and support data-driven modeling tasks. By producing high-quality data, they facilitate the development of machine learning-based thermal and energy models and enable the creation of computationally efficient black- and grey-box models for all prototype buildings in the PNNL buildings database. These contributions offer significant advancements in building energy simulation and modeling, paving the way for improved research and practical applications in the field.

\section*{Acknowledgment}
This work is funded by the NSF award 2208783.

\bibliographystyle{IEEEtran}

\begin{thebibliography}{10}
\providecommand{\url}[1]{#1}
\csname url@samestyle\endcsname
\providecommand{\newblock}{\relax}
\providecommand{\bibinfo}[2]{#2}
\providecommand{\BIBentrySTDinterwordspacing}{\spaceskip=0pt\relax}
\providecommand{\BIBentryALTinterwordstretchfactor}{4}
\providecommand{\BIBentryALTinterwordspacing}{\spaceskip=\fontdimen2\font plus
\BIBentryALTinterwordstretchfactor\fontdimen3\font minus \fontdimen4\font\relax}
\providecommand{\BIBforeignlanguage}[2]{{%
\expandafter\ifx\csname l@#1\endcsname\relax
\typeout{** WARNING: IEEEtran.bst: No hyphenation pattern has been}%
\typeout{** loaded for the language `#1'. Using the pattern for}%
\typeout{** the default language instead.}%
\else
\language=\csname l@#1\endcsname
\fi
#2}}
\providecommand{\BIBdecl}{\relax}
\BIBdecl

\bibitem{EIAEnergyConsumptionWebpage}
\BIBentryALTinterwordspacing
``Consumption \& efficiency - u.s. energy information administration (eia),'' "Accessed: 2024-11-18". [Online]. Available: \url{https://www.eia.gov/consumption}
\BIBentrySTDinterwordspacing

\bibitem{xiang2022historical}
X.~Xiang, M.~Ma, X.~Ma, L.~Chen, W.~Cai, W.~Feng, and Z.~Ma, ``Historical decarbonization of global commercial building operations in the 21st century,'' \emph{Applied Energy}, vol. 322, p. 119401, 2022.

\bibitem{zhou2016achieving}
Z.~Zhou, S.~Zhang, C.~Wang, J.~Zuo, Q.~He, and R.~Rameezdeen, ``Achieving energy efficient buildings via retrofitting of existing buildings: a case study,'' \emph{Journal of Cleaner Production}, vol. 112, pp. 3605--3615, 2016.

\bibitem{chen2018measures}
Y.~Chen, P.~Xu, J.~Gu, F.~Schmidt, and W.~Li, ``Measures to improve energy demand flexibility in buildings for demand response (dr): A review,'' \emph{Energy and buildings}, vol. 177, pp. 125--139, 2018.

\bibitem{shan2016building}
K.~Shan, S.~Wang, C.~Yan, and F.~Xiao, ``Building demand response and control methods for smart grids: A review,'' \emph{Science and Technology for the Built Environment}, vol.~22, no.~6, pp. 692--704, 2016.

\bibitem{gunay2019data}
H.~B. Gunay, W.~Shen, and G.~Newsham, ``Data analytics to improve building performance: A critical review,'' \emph{Automation in Construction}, vol.~97, pp. 96--109, 2019.

\bibitem{EnergyPlusToolsDirectory}
\BIBentryALTinterwordspacing
``Best directory: Tools built on energyplus,'' accessed: 2024-11-18. [Online]. Available: \url{https://www.ibpsa.us/best-directory-list/}
\BIBentrySTDinterwordspacing

\bibitem{alanne2022overview}
K.~Alanne and S.~Sierla, ``An overview of machine learning applications for smart buildings,'' \emph{Sustainable Cities and Society}, vol.~76, p. 103445, 2022.

\bibitem{OpenStudioDocumentation}
\BIBentryALTinterwordspacing
``Openstudio documentation,'' accessed: 2024-11-18. [Online]. Available: \url{https://openstudio.net/}
\BIBentrySTDinterwordspacing

\bibitem{CityBESDocumentation}
\BIBentryALTinterwordspacing
``Citybes documentation,'' accessed: 2024-11-18. [Online]. Available: \url{https://buildings.lbl.gov/urban-science/tools}
\BIBentrySTDinterwordspacing

\bibitem{present2024resstock}
E.~Present, P.~R. White, C.~Harris, R.~Adhikari, Y.~Lou, L.~Liu, A.~Fontanini, C.~Moreno, J.~Robertson, and J.~Maguire, ``Resstock dataset 2024.1 documentation,'' National Renewable Energy Laboratory (NREL), Golden, CO (United States), Tech. Rep., 2024.

\bibitem{parker2023comstock}
A.~Parker, H.~Horsey, M.~Dahlhausen, M.~Praprost, C.~CaraDonna, A.~LeBar, and L.~Klun, ``Comstock reference documentation (v. 1),'' National Renewable Energy Laboratory (NREL), Golden, CO (United States), Tech. Rep., 2023.

\bibitem{EnergyPlus}
\BIBentryALTinterwordspacing
 [Online]. Available: \url{https://energyplus.net/}
\BIBentrySTDinterwordspacing

\end{thebibliography}

\end{document}